\documentclass[useAMS,usenatbib]{mn2e}

\usepackage{graphicx}

\title[Far Infrared Luminosity Function of Local Star-forming Galaxies]{Far Infrared Luminosity Function of Local Star-forming Galaxies in the AKARI Deep Field South}
\author[C. Sedgwick et al]
{Chris Sedgwick$^{1}$, 
Stephen Serjeant$^{1}$,
Chris Pearson$^{4,2,1}$,
 Shuji Matsuura$^{3}$,
  \newauthor
Mai Shirahata$^{3}$, 
Shinki Oyabu$^{8}$,
Tomotsugu Goto$^{5,6}$,
Hideo Matsuhara$^{3}$,
\newauthor
D. L. Clements$^{7}$,
Mattia Negrello$^{1}$,
and Glenn J. White$^{1,2}$\\
$^{1}$Department of Physics \& Astronomy, The Open University, Milton Keynes MK7 6AA\\
$^{2}$Rutherford Appleton Laboratory, Chilton, Didcot, Oxfordshire OX11 0QX\\
$^{3}$Institute of Space and Astronautical Science, Japan Aerospace Exploration Agency, Sagamihara, Kanagawa, 229 8510, Japan\\
$^{4}$Institute for Space Imaging Science, University of Lethbridge, Lethbridge, Alberta, T1K 3M4, Canada\\
$^{5}$Institute for Astronomy, University of Hawaii, 2680 Woodlawn Drive, Honolulu, HI 96822, USA\\
$^{6}$Subaru Telescope, 650 North A'ohoku Place, Hilo, HI 96720, USA\\
$^{7}$Astrophysics Group, Imperial College, Blackett Laboratory, Prince Consort Road, London SW7 2AZ\\
$^{8}$Graduate School of Science, Nagoya University, Furo-cho, Chikusa-ku, Nagoya, Aichi 464-8602, Japan}

\begin{document}

\date{Revision submitted May 2011}

\pagerange{\pageref{firstpage}--\pageref{lastpage}} \pubyear{2011}

\maketitle

\label{firstpage}

\begin{abstract}
We present a far-infrared galaxy luminosity function for the local universe. We have obtained 389 spectroscopic redshifts for galaxies observed at 90 $\mu$m  in the AKARI Deep Field South, using the AAOmega fibre spectrograph via optical identifications in the digitized sky survey and 4m-class optical imaging. For the luminosity function presented in this paper, we have used those galaxies which have redshifts $0<z<0.25$, have optical magnitudes and are not part of a newly discovered cluster of galaxies (giving a total of 130 sources). Infrared and optical completeness functions were estimated using earlier Spitzer data and APM B-band optical data respectively, and the luminosity function has been prepared using  the $1/V_{\rm max}$ method. We also separate the luminosity function between galaxies which show evidence of predominantly star-forming activity and predominantly active galactic nucleus (AGN) activity in their optical spectra. Our luminosity function is in good agreement with the previous 90 $\mu$m luminosity function from the European Large Area ISO Survey, and we also present a luminosity function with combined AKARI and ISO data. The result is in reasonable agreement with predictions based on the earlier IRAS $60\,\mu$m PSC-z catalogue, and with a recent backward evolution model.\end{abstract}


\section{Introduction}

Recent advances in observational infrared astronomy have led to a rapid increase in the constraints which can now be placed on theoretical models of the development of structure in the Universe. The discovery of far-infrared luminous populations of galaxies by SCUBA (e.g. Hughes et al. 1998, Smail et al. 1997) and of dust obscured galaxies by Spitzer (Dey et al. 2008) have led to the realisation that massive galaxies formed the bulk of their stars at high redshifts during bursts of star-formation, contrary to the simpler growth expectations predicted by earlier hierarchical models (see review in Hauser \& Dwek 2001). Furthermore, there has been a growing realisation that feedback from intermittent AGN activity is important in regulating stellar mass assembly (e.g. di Matteo et al. 2005). 

The evolution of  star-forming galaxies (SFGs) and galaxies with active galactic nuclei (AGNs) can provide important observational constraints on galaxy evolution models. Radiation from AGN dust tori is predicted to peak in the mid-infrared (Efstathiou \& Rowan-Robinson 1995), whereas dust radiation from the giant molecular clouds which dominate the bolometric output from star formation means that SFGs will peak in the far-infrared, between 50 $\mu$m - 200 $\mu$m.

There has been considerable work recently in the infrared shortward of this SFG peak. A number of key papers on the history of star formation and the evolution of the  luminosity function have used  24 $\mu$m data from the Spitzer Space Telescope. An extensive study of the evolution of star-forming galaxies showed an order of magnitude increase in typical infrared luminosity from the present out to  $z=2$ (P\'{e}rez-Gonz\'{a}lez et al. 2005). A recent study of Spitzer  8$\,\mu$m - 24$\,\mu$m sources suggested a rapid increase of the co-moving volume emissivity back to $z$ $\sim1$ and a constant average emissivity at $z>1$ as well as showing evidence of downsizing in star formation (Rodighiero et al. 2010). Magnelli et al. (2009) presented luminosity functions at 15 $\mu$m (based on Spitzer 24 $\mu$m data) and at 35 $\mu$m (based on Spitzer 70 $\mu$m data) between redshifts $0.4< z < 1.3$. Rujopakarn et al. (2010) presented a 24 $\mu$m luminosity function in the range $0< z<0.65$, and also combined their results with those of Magnelli et al. (2009) to cover redshifts out to $z=1.2$, confirming strong evolution, consistent with pure luminosity evolution. Several studies have found that star formation from Luminous InfraRed Galaxies (LIRGs, 10$^{11}$L$_{\odot}$$<$L$_{IR}$$<$10$^{12}$L$_{\odot}$) and Ultra-Luminous InfraRed Galaxies (ULIRGs, 10$^{12}$L$_{\odot}$$<$L$_{IR}$$<$10$^{13}$L$_{\odot}$) have made a much larger fractional contribution to the co-moving volume-averaged star formation density at $z=1$ than they do locally (e.g. Le Floc'h et al. 2005, Caputi et al. 2007). A study of data from the AKARI Space Telescope's Deep Field North based on observed wavelengths of 2$\,\mu$m - 24$\,\mu$m which analysed data out to $z\sim2.2$ has also found strong evidence for evolution of galaxy luminosities towards higher redshift (Goto et al. 2010).

Work has also been done recently at  wavelengths longer than the SFG peak. A  measurement of the ISO 170 $\mu$m luminosity function was made by Takeuchi et al. 2006. Luminosity functions based on observations at 250, 350 and 500\,$\mu$m from the recent Balloon-borne Large Aperture Submillimeter Telescope (BLAST) survey centred on the GOODS-South field have shown a strong increase in the number density of the most luminous galaxies from the present out to $z=1$ (Eales et al. 2009). Two recent papers have presented luminosity functions using observations at 250 $\mu$m from the science demonstration phase of the Herschel Space Telescope: Dye et al. (2010) used data from a 14 deg$^2$ field out to $z=0.5$, showing steady evolution out to this redshift, and Eales et al. (2010) in two smaller but deeper fields to $z=2.0$ found strong evolution out to $z\sim1$ but at most weak evolution beyond this redshift.  

The first measurements at wavelengths closer to the SED peaks of local star-forming galaxies were made by the Infra-Red Astronomical Satellite (IRAS). Saunders et al. (1990) presented a local luminosity function of 60 $\mu$m-selected IRAS sources and found evidence of strong evolution (which could be explained by evolution in either luminosity or density). This work was extended and revised by Soifer \& Neugebauer (1991), Sanders et al. (2003) and Takeuchi et al. (2003). Wang \& Rowan-Robinson (2010) confirmed this evolution using a much larger sample of IRAS 60 $\mu$m sources in the redshift range $0.02<z<0.1$. The evidence from the ISO ELAIS 90 $\mu$m survey suggested a calibration disagreement with the IRAS 100 $\mu$m data (H\'{e}raudeau et al. 2004) so it is important to obtain independent estimates of the local luminosity function, particularly at 90 - 100 $\mu$m.  A 90$\,\mu$m luminosity function was derived from the European Large Area ISO Survey (ELAIS paper IX,  Serjeant et al. 2004), based on spectroscopic redshifts of 151 Infrared Space Observatory (ISO) sources in the Northern ELAIS fields. With optical identification of 29 ISO far-infrared (FIR) sources in the Lockman Hole using the FIR-radio correlation for positional accuracy, Oyabu et al. (2005) showed that the space density of the FIR sources at $0.3< z<0.6$ was some 460 times larger than the local value. A recent luminosity function using Spitzer 70 $\,\mu$m sources in the ultra-deep GOODS-North pencil-beam survey showed evidence for evolution at redshifts $z>0.4$ (Huynh et al. 2007). An analysis to find the split between SFGs and AGN-powered galaxies in Spitzer 70$\,\mu$m data in the Extended Groth Strip field found that virtually all were powered predominantly by star-forming activity (Symeonidis et al. 2010).

AKARI's Far Infrared Surveyor Instrument (FIS) is most sensitive at 90$\,\mu$m, and provides the opportunity to expand the observational evidence in this region near the SFG peak. This paper aims to provide a new 90$\,\mu$m luminosity function, to compare with the previous ISO one, and also to combine with data from that survey to provide the best luminosity function at this wavelength so far. The present study aims to measure a local benchmark against which to compare luminosity functions of more distant SFGs based on more detailed, deeper data which will become available in the near future from instruments such as Herschel Space Telescope. In this paper, we will constrain the 90$\,\mu$m galaxy luminosity function for $z<0.25$, using the spectroscopic redshifts we have found with the AAOmega fibre spectroscope for sources in the AKARI Deep Field South (ADF-S, Matsuura et al., 2011). An earlier version of this luminosity function (using about one-half  of the data now available, showing similar results) was included as part of a conference proceedings (Sedgwick et al. 2009). In a separate paper (Goto et al. 2011) we are estimating the local far infrared luminosity function by cross-correlating the AKARI all-sky survey with IRAS Bright Source catalogue. 

The methods used to find the infrared sources, to identify redshifts from cross-identified optical spectra of these sources and to derive the luminosity function, are described in Section \ref{sec:methods}. The results are presented in Section \ref{sec:results} and discussed in Section \ref{sec:discussion}. 

This paper assumes the parameter values of the current concordance cosmology: $H_0$=72.0 km s$^-$$^1$Mpc$^-$$^1$, $\Omega_M$=0.3 and $\Omega_{\Lambda}$=0.7.

\section[]{Methods}\label{sec:methods}

\subsection{AKARI Deep Field South Observations}

The AKARI Space Telescope (Murakami et al. 2007) is a cryogenically cooled 68.5 cm mid/far-infrared
telescope in a low Earth, Sun-synchronous orbit. Besides performing an
all-sky survey at $10\,\mu$m-$160\,\mu$m, AKARI also conducted a series of
deep pointed observations. These deep guaranteed time extragalactic survey
programmes were conducted near the North and South Ecliptic Poles for
reasons of orbital visibility, and are described in Matsuhara et al.
(2006).  Both the FIS ($60-160\,\mu$m) and the InfraRed Camera
(IRC, $2-24\,\mu$m) instruments were used in these surveys. 

The AKARI Deep Field South (ADF-S) is centered on RA 4h 44m 00s, 
Dec $-53$\degr ~20\arcmin~00\arcsec ~ (J2000), a low-cirrus region near the South Ecliptic Pole. The FIS survey in this region covered over 12  deg$^2$. IRC imaging was also conducted in Open Time over part of this field (about 0.8 deg$^2$). 

The sources reported in this paper were detected in the ADF-S by the 90$\,\mu$m WIDE-S band of
the FIS, using its slow-scan observation mode with point-source photometry (Shirahata et al. 2009). This reached $28\,$mJy at  $60\%$ completeness (our infrared completeness function is based on earlier Spitzer data: see Section \ref{sec:lf_methods} for details). By way of comparison, a 50\% completeness limit of the sources used in  the 90$\,\mu$m luminosity function for the ELAIS Northern fields reached 70 mJy (Serjeant et al. 2004).

The FIS survey found a total of 2,282 sources at $90\,\mu$m at a signal-to-noise ratio SNR$>$5 (giving a minimum flux of 12.81 mJy) in this field (Shirahata et al., in preparation). Cross-identifications with other infrared catalogues are shown in Table~\ref{table:xids}.

\begin{table}
\caption{Cross-identifications between the AKARI FIS 90 $\mu$m catalogue and other infrared catalogues in the ADF-S. The 1$\sigma$ positional uncertainty shown below is based on the formula $\delta$RA=$\delta$DEC=0.6(1/SNR)FWHM (Ivison et al. 2007) for FIS and Spitzer. The BLAST uncertainty is quoted in Valiante et al. (2010). Therefore a 2$\sigma$ radius for a combination of sources is between about 10" and 15", so matches at these distances are shown below. For AKARI FIS details (data not yet published), see Shirahata et al. (2009). For BLAST data and results, see Valiante et al. (2010). For Spitzer results (data not yet published), see Scott et al. (2010). AKARI IRC data has not yet been published.}\label{table:xids}
\begin{center}
\begin{tabular}{|ccccc|}
\hline

 &  &  & \multicolumn{2}{c|}{Matches to }  \\ 
 &  &  & \multicolumn{2}{c|}{FIS 90 $\mu$m}  \\ 
                  &    Total    &1$\sigma$  & \multicolumn{2}{c|}{ sources within}  \\   
Catalogue   & Sources & Error &  10\arcsec & 15\arcsec~~ \\   
 \hline
AKARI FIS 90  $\mu$m          &    ~2,282      &  5.1\arcsec    &                      &     \\
AKARI FIS 65 $\mu$m           &   ~~~ 391     &  5.4\arcsec    &     ~~125     &      ~ 174 \\
AKARI FIS 140 $\mu$m         &    ~~~315     &  7.0\arcsec   &      ~~ 56      &    ~~~80  \\
AKARI FIS 160 $\mu$m         &     ~~~216    &  6.5\arcsec   &     ~~~~7      &    ~~~11  \\
Spitzer 70 $\mu$m                  &    ~~1,205     &  5.2\arcsec  &      ~~583     &         ~72  \\
Spitzer 24 $\mu$m                  &     41,503       &  1.1\arcsec  &      1,199      &     1,584  \\
AKARI IRC 15  $\mu$m         &      ~~~725     &   2.0\arcsec  &   ~~~25      &    ~~~28  \\
AKARI IRC 24  $\mu$m         &       ~~~184    &  2.0\arcsec   &    ~~~24      &   ~~~30  \\
BLAST 250/350/500 $\mu$m  &   ~~~232     &  5.0\arcsec &      ~~~22      &   ~~~38  \\

\hline
\end{tabular}
\end{center}
\label{default}
\end{table}

\subsection{Optical Identifications}

We used the Bj-band APM catalogue for optical cross-matching of FIS sources, and for estimating the optical completeness of the FIS catalogue (discussed in Section~\ref{sec:lf_methods}). This allows a comparison with earlier ELAIS work, which is in the Bj-band.

Optical identifications of the AKARI sources were checked by eyeballing all the candidate sources using R-band images from the Digitised Sky Survey (DSS) and (where available) images from the CTIO 4m telescope. We chose DSS-R for eyeballing identifications of both IRC (mid-infrared) and FIS sources, partly because this better suited the IRC population and partly to match the available CTIO R-band data over part of the field. Where available, identifications were supported by sub-arcsecond positions from the $1.4$\,GHz Australia Telescope Compact Array survey (White et al. in preparation) and the AKARI-IRC data (Pearson et al. in preparation). Since we are dealing in this paper with bright local FIS (far-infrared) galaxies, identifications in DSS B-band have counterparts in the DSS-R band, and vice versa.


\subsection{AAOmega Spectroscopic Redshift Identifications}

We used AAOmega, the fibre-fed optical spectrograph at the  Anglo
Australian Observatory, to obtain optical spectra of a selection of
the ADF-S sources in order to determine spectroscopic
redshifts. For our targets, we used a sparse sample based on the AAOmega fibre configuration, which (in the absence of any evidence to the contrary) we have assumed to be unbiased. The area covered was $\pi$ degrees (one AAOmega field-of-view: see Figure~\ref{fig:field}). Only sources for which we had optical counterparts were targeted, and these were cross-identified with the AKARI
source coordinates by eye. A few very local sources were extended optical objects,
 and the cross-identified AKARI coordinates were within the extended optical image.
Most of the cross-identified optical sources were point sources, 94\% of which fell
within 10\arcsec, and all fell within 16\arcsec  of the
AKARI coordinates. The AKARI telescope pointing error is less than 3\arcsec (Murakami et al. 2007).

Data were obtained by AAOmega during four sessions between
October 2007 and November 2008. None of the nights was
photometric. Gratings were 385R and 580V with central wavelengths 7250\AA \thinspace  
and 4800\AA, and wavelength resolution $\lambda$/$\Delta$$\lambda$=1300. The spectra were reduced using the standard AAOmega pipeline, which was also used to combine the red and blue arms. The combined spectra each have a range of about 3800 - 8800\AA, although noise at either end of this range, particularly the blue end, often reduced the useful data to a range of about 4000 - 8700\AA.

\begin{figure}
  \resizebox{\hsize}{!}{\includegraphics{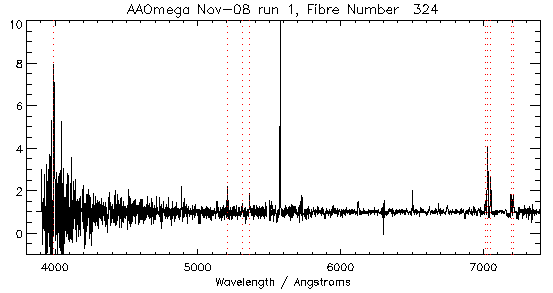}}
 \resizebox{\hsize}{!}{\includegraphics{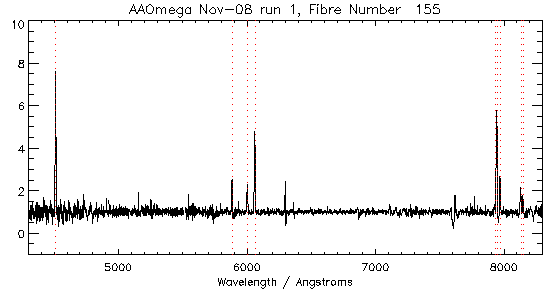}}
\caption{Examples of spectra which show the main emission lines used to identify redshifts in this paper:  from left, [OII]$\lambda3726$, H$\beta$/[OIII]/[OIII] $\lambda\lambda4861, 4958, 5007$, [NII]/H$\alpha$/[NII] $\lambda\lambda6548, 6563, 6583$ and [SII]$\lambda\lambda6716, 6731$ (rest-frame wavelengths). Note that the strong emission line at $\sim$5570\AA~is an artefact from street lighting in a local town; there are also some telluric lines (e.g. at $\sim$7700\AA). The spectra have been normalised to 1 using boxcar median smoothing to remove the effect of differing throughputs by the blue and red arms of the spectrograph. Using the Kauffmann criterion on the BPT diagram as described in the text, the top source (which is at $z=0.071$) is an SFG, and the bottom source ($z=0.211$) is an AGN.}\label{fig:spectra}
\end{figure}

\begin{figure}
  \resizebox{\hsize}{!}{\includegraphics{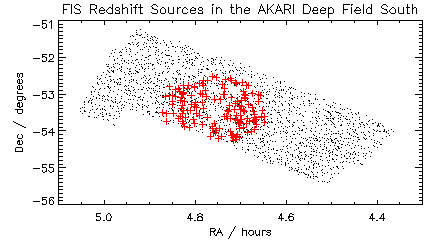}}
\caption{FIS sources with z$<$0.25 (red plus symbols) on all FIS sources (black dots) in the AKARI Deep Field South.}\label{fig:field}
\end{figure}

Redshifts were measured using the following emission lines: [NII]/H$\alpha$/[NII] $\lambda\lambda$6548, 6563, 6583 and the [SII] doublet $\lambda\lambda$6716, 6731, and the three H$\beta$/[OIII]/[OIII] $\lambda\lambda$4861, 4958, 5007 emission lines. Redshifts were only taken where one or other (or both) of these two triplets of emission lines were identified. Unless it was redshifted out of range, the [SII] pair was invariably seen with the [NII]/H$\alpha$/[NII] triplet. In addition, the asymmetrical [OII] emission line resulting from the close doublet $\lambda\lambda$ 3726, 3729  was often found at the same redshift to support the identifications. Examples of spectra obtained showing the redshift identifications are given in Figure~\ref{fig:spectra}.

\begin{table}
\caption{AAOmega observations: exposure times and number of redshifts identified from emission lines.}\label{table:exposures}
\begin{center}
\begin{tabular}{|ccc|}
\hline

Date                 & Total Exposure      & Redshifts  \\ 
                          & hr:m:s     & Identified \\  
 \hline
29 Oct 2007         &    2:30:00      &    130    \\
30 Nov 2007       &    1:23:40      &     ~63    \\
3 Jan 2008          &    2:00:00      &     ~59   \\
27 Nov 08 a         &    1:46:40     &     ~76   \\
27 Nov 08 b         &    0:26:40     &    ~61    \\
Total                      &                       &   389  \\

\hline
\end{tabular}
\end{center}
\end{table}

The quality of the spectra obtained varied widely between the different nights, as a result of varying asmospheric conditions and length of exposure times achieved for each set of targets (see Table~\ref{table:exposures}).  Overall, a total of 389 redshifts were measured for the 1179 distinct sources targeted. Some of these were for non-FIS sources (e.g. AKARI-IRC, BLAST) and will be included in a future paper. For the FIS sources used in this study, a total of 317 sources for which we obtained optical counterparts were targeted, and redshifts were obtained for 183, a success rate of about 58\%. Of these, 152 had redshifts in the range $0<z<0.25$ and also had optical magnitudes, and these (subject to the removal of a new cluster  described in \ref{sec:cluster}) will be used in preparing the luminosity function presented in this paper, and are shown in Figure~\ref{fig:field}. Sources without spectroscopic redshifts were not used.

\begin{figure}
  \resizebox{\hsize}{!}{\includegraphics{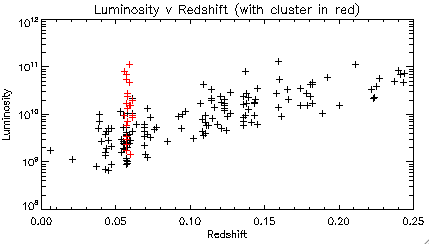}}
\caption{The redshift - luminosity plane for all sources used, plus the cluster sources (which are highlighted in red). Luminosity is $\nu$L$_\nu$/L$_{\odot}$ as calculated for the luminosity functions in this paper.} \label{fig:redshift_lum}
\end{figure}

The distribution of luminosity against spectroscopic redshifts used in preparing this
luminosity function  is shown in Figure~\ref{fig:redshift_lum} (which also highlights the new cluster found at $z\sim0.06$ discussed in Section \ref{sec:cluster} below).

\subsection{AGN / SFG Discrimination}

Two indicators were used to discriminate between AGNs and
SFGs in our sources.

(a) Quasars are often defined as sources which have broad Balmer emission lines
with FWHM over 1000 km s$^-$$^1$. Although we were able to measure FWHM of the H$\alpha$ line for all the 152 selected sources, only three of them had such broad
emission lines. In this paper, we follow Vanden Berk et al. (2001), and
regard sources with  FWHM of the H$\alpha$ line over 500 km s$^-$$^1$
as indicating significant AGN activity. A total of 10 of the sources in our sample
showed this. 
 
\begin{figure}
 \resizebox{\hsize}{!}{\includegraphics{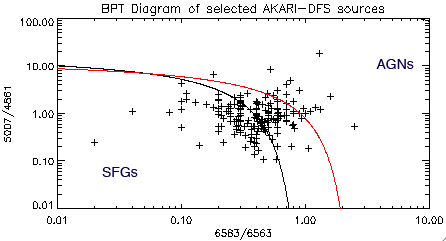}}
\caption{BPT diagram of the 105 sources used in the luminosity functions which had data to calculate these two ratios. The black line plots the Kauffmann et al. (2003) formula to discriminate between AGNs and SFGs. The area between this black line and the red line contains AGN-SFG composite sources, based on the formulae in Kewley et al. (2006).}\label{fig:bpt}
\end{figure}

 (b) A BPT diagram (Baldwin, Phillips \& Terlevich 1981) is plotted in Figure~\ref{fig:bpt},  relating the two intensity ratios of [NII] $\lambda$6583 / H$\alpha$ $\lambda$6563 and of  [OIII] $\lambda$5007 / H$\beta$ $\lambda$4861. The black line discriminates between AGN sources and SFG sources, based on the formula in Kauffmann et al. (2003), with a position above this line indicating an AGN. A total of 105 sources had sufficient spectral data to plot the ratios in this diagram, of which 39 were identified as AGNs by the Kauffmann line.

Most of the sources classified as AGNs were therefore selected by the BPT method, although many of these were not broad-line AGN since they had FWHM of the H$\alpha$ line  $<$ 500 km s$^{-1}$. In addition, there were 5 sources with the H$\alpha$ line $>$ 500 km s$^-$$^1$,  3 of which did not have the ratios to use the BPT diagram, and 2 of which had ratios below the BPT Kauffmann classifying line. Classifying these additional sources as AGNs did not significantly affect the separate luminosity functions in Section \ref{sec:results} below.
 
The ratio of  [OIII]  $\lambda$5007 /[OII]  $\lambda$3727 was also considered as a possible AGN/SFG discriminator. However, this ratio proved unreliable: the two emission lines are not close and are from the two (blue and red) arms of the spectrograph which have different throughputs; in addition, there could be reddening intrinsic to the galaxies. In fact, this indicator usually conflicted sharply with the other indicators. In contrast, the ratios used in the BPT diagram are of lines which are close in wavelength, so are not affected either by throughput variations or by reddening.

The result of this analysis gave 44 optically-selected AGN sources and 108 optically-selected SFG sources out of the total of 152 FIS sources with $z<0.25$ and optical magnitudes. Note, however, that many of our sources lie close to the dividing line, and the Kewley et al. (2006) criterion suggests many are AGN-SFG composite sources (Figure \ref{fig:bpt}). Both star formation and active nuclei contribute to the bolometric energy output of these galaxies, although the BPT diagram on its own is insufficient to determine bolometric fractions.

 \begin{figure}
 \resizebox{\hsize}{!}{\includegraphics{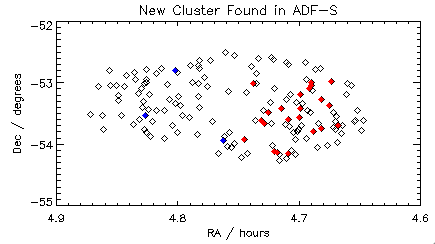}}
\caption{The new cluster found in the survey field. Cluster members are shown as filled red diamonds; sources in the same redshift range but not cluster members as filled blue diamonds; other redshift sources as open diamonds.}\label{fig:cluster}
\end{figure}

\subsection{Cluster of Galaxies in the Field}\label{sec:cluster}

As can be seen in the  luminosity / redshift plot (Figure~\ref{fig:redshift_lum}), there is a group of sources bunched at about $z=0.06$. In fact, there are 22 sources at a redshift of between $0.056<z<0.061$ and a separation on the sky of under 45 arcminutes, centred on  (04 42 12.5, -53 30 46). This redshift range corresponds to a depth of 21 Mpc, and this angular separation corresponds to a width of about 3.2 Mpc. The cluster sources are identified in Figures~\ref{fig:redshift_lum} and \ref{fig:cluster} and listed in Table~\ref{table:cluster}. Figure~\ref{fig:redshift_lum} shows that the cluster contains relatively more luminous sources than the sample as a whole. This may have been caused by under-representation of high-luminosity galaxies at the lowest redshifts in our sample, or we may have sampled a region of enhanced star-formation in the outer (infalling) region of the cluster (Coppin et al. 2011, but see Bai et al. 2009 and Haines et al. 2011).  Including this cluster in the sample distorts the luminosity function and causes the mean of V/$V_{\rm max}$ to fall well below 0.5 (this is discussed in Section \ref{subsec:VVmax} below). Therefore, we have excluded the sources in the cluster from the final luminosity function.  (This is not the well-known Dressler DC0428-53 cluster at $z=0.041$, which is within the ADF-S field but outside the part of the field for which we have obtained redshifts.)

We have estimated the comoving volume of the cluster as $\sim$10$^4$ Mpc$^3$. For this calculation, we  converted the redshift range to comoving depth per unit area and multiplied by the area of the cluster. This cluster volume is a tiny proportion ($\sim0.01$) of the whole field, so excluding the cluster volume from the calculations makes no significant difference to either the luminosity function or $\langle$V/V{\tiny max}$\rangle$ so we have not made this adjustment.\\

\subsection{Luminosity Function Methods}\label{sec:lf_methods}

The 90 $\mu$m local luminosity function was estimated as follows. 

\begin{figure}
 \resizebox{\hsize}{!}{\includegraphics{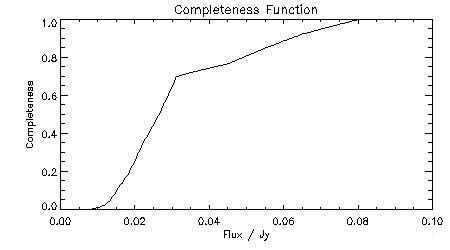}}
\caption{Estimated FIS completeness function using Lagache model (Lagache et al. 2004) fitted to Spitzer 70 $\mu$m GOODS-N data, modified for the 90 $\mu$m case by the Pearson model (Pearson 2001) as described in the text.}\label{fig:completeness}
\end{figure}

Completeness as a function of flux based directly on the FIS data is still under construction (Shirahata et al., in preparation). Therefore we have had to make our own assessment of the completeness of the AKARI data. The ultra-deep Spitzer 70 $\mu$m observations of the GOODS-North field reached down to a flux of 1.2 mJy (Frayer et al. 2006). This is an order of magnitude lower than the limit of the present AKARI survey, so we assume this will be 100\% complete down to the lower end of the AKARI data. Frayer et al. (2006) showed that this Spitzer data is well fitted by the model of Lagache et al. (2004) which we used to generate number counts against which to measure the AKARI completeness. However, since this data is at 70 $\mu$m we have had to make an adjustment for our 90 $\mu$m case, and to do this we used the model described in Pearson (2001), scaling the fluxes by a factor of 0.673 on the basis of that model's prediction that $N(S_{70\,\mu{\rm m}}>0.0673{\rm\,Jy})=N(S_{90\,\mu{\rm m}}>0.1{\rm\,Jy}$). The resulting completeness ranges from  97\% at 100 mJy to
60\% at 28 mJy, as shown in Figure~\ref{fig:completeness}.

\begin{figure}
 \resizebox{\hsize}{!}{\includegraphics{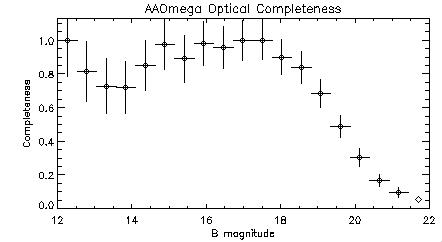}}
\caption{Estimated AAOmega optical completeness function using APM catalogue. Vertical lines show 1$\sigma$ errors, and horizontal lines show bin sizes.}\label{fig:opt_completeness}
\end{figure}

As mentioned earlier, sources without APM B-band counterparts were excluded. Using APM B-magnitudes, we estimated spectroscopic completeness as a function
of magnitude, using the AAOmega redshift sources compared to APM sources cross-identified with the sources in the total ADF-S FIS sample before identification of redshifts.  Each point in Figure~\ref{fig:opt_completeness} shows the completeness using a range of $\pm$1.5 magnitudes. The 1$\sigma$ error bars of the points are shown in the figure. The completeness is close to 100\% down to B-band magnitude of about 18 after which it declines to 0.1 at about magnitude 21.

For galaxies with B$>$21 and z$<$0.25, we have adopted the assumption used for ELAIS (see appendix to Serjeant et al. 2001 for a detailed discussion) that such galaxies have more local counterparts, and at least one member of such a class of galaxy would be included in our sample. This would mean that its contribution to the luminosity function using the 1/V{\tiny max} method would be given a weighting that exactly accounts for the increased difficulty of detecting such a class of galaxy.

To check whether or not our sample is typical of the sky as a whole - i.e. whether it is affected by large scale cosmic variance -  we compared the number per unit area with that obtained in the 50 deg$^2$ Spitzer Wide-area Infrared Extragalactic Legacy Survey (SWIRE) at 70 $\mu$m (quoted in Frayer et al. 2009)  for $>0.1$\,Jy objects  (following the procedure adopted in ELAIS, e.g. Serjeant et al. 2001). The fraction of number predicted / actual number was 1.80$\pm0.32$ and we renormalised the effective luminosity density (and error) by this fraction.
 
\begin{figure}
 \resizebox{\hsize}{!}{\includegraphics{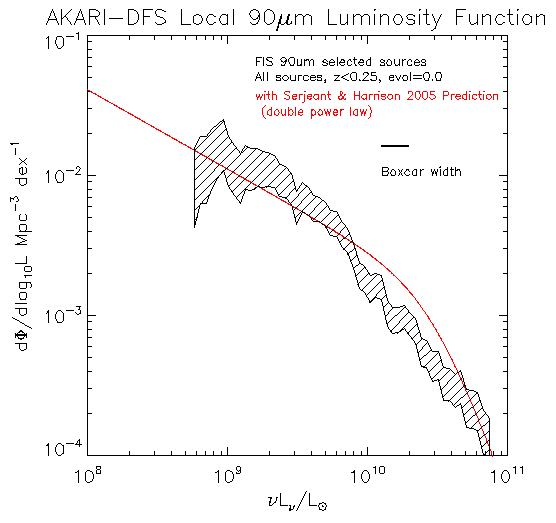}}
\caption{Local galaxy luminosity function using all ADF-S sources with AAOmega spectroscopic redshifts selected as described in the text. It is assumed in this figure that there is no pure luminosity evolution. The hatched area shows Poisson error bands. The line is the double power-law prediction by Serjeant \& Harrison 2005.}\label{fig:all_lf}
\end{figure}

As explained above, we have ended up with a final total of 130 sources (after excluding the cluster) for the luminosity function. For comparison, the fields used in the ELAIS IX luminosity function had a
total of 151 sources (as with our sample, these all had spectroscopic redshifts). More sources would obviously have been welcome, and we hope to obtain further AAOmega observations in the future. This survey covers a similar redshift range to the ELAIS IX survey (cf. Figure 1 of Serjeant et al. 2004), so in addition to showing our new results (Figures~\ref{fig:all_lf} to \ref{fig:agn_lf}), we have felt justified in combining the two samples. The combined luminosity function is shown in Figure~\ref{fig:combined_lf} and discussed below.
 
We calculated the 1/V{\tiny max} luminosity function (Schmidt 1968) following the methodology in Serjeant et al. (2004). K-corrections were made assuming the M82 star-forming spectral energy distribution, except for the AGN luminosity function for which we assumed a Seyfert 2 SED template derived from the Spitzer Wide-Area Infrared Extragalactic Survey (SWIRE).

\section{Results}\label{sec:results}

\subsection{ADF-S Results}\label{subsec:results_akari}

As expected from earlier studies (Le Floc'h et al. 2005, Caputi et al. 2007), our $z<0.25$ sample is largely free of LIRGs and ULIRGs. The highest luminosity was 1.3 x 10$^1$$^1$ L$_{\odot}$, and only two sources were over 10$^1$$^1$L$_{\odot}$,  the usual low-end definition of LIRGs (Sanders \& Mirabel 1996), where we use $\nu$L$_{\nu}$(90$\mu$m) as a rough estimate of the bolometric infrared luminosity.

The local 90 $\mu$m luminosity function for all
sources is shown in Figure~\ref{fig:all_lf}, and tabulated in Table~\ref{table:tab}. For each luminosity bin, we have calculated the space density of objects with
luminosities within $\pm$0.15 dex. 

\begin{figure}
\begin{center}
 \resizebox{2.6in}{!}{\includegraphics{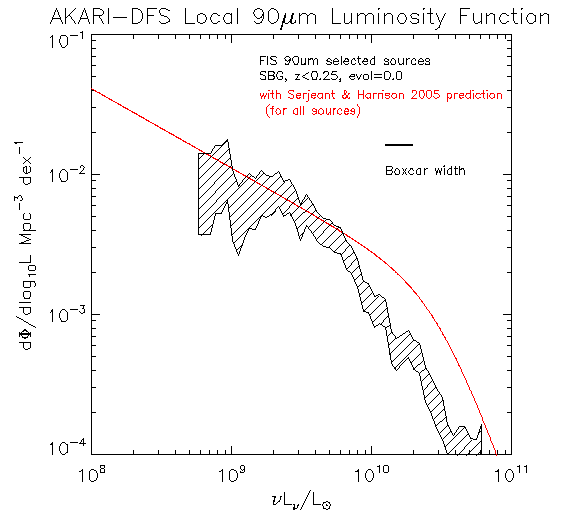}}
\caption{Star-forming galaxy luminosity function (with AGNs excluded). The line is the double power-law prediction by Serjeant \& Harrison 2005 (for all galaxies).}\label{fig:sbg_lf}
\end{center}
\end{figure}

\begin{figure}
\begin{center}
 \resizebox{2.6in}{!}{\includegraphics{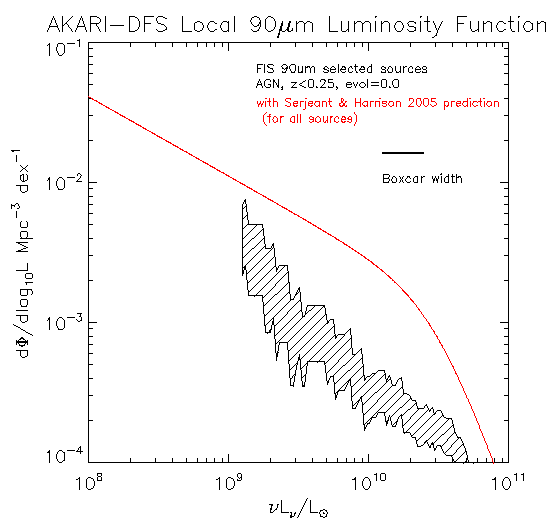}}
\caption{AGN luminosity function (with SFGs excluded) as described in the text. The line is the double power-law prediction by Serjeant \& Harrison 2005 (for all galaxies). Note that only 36 AGNs were identified in the sample of 130 sources.}\label{fig:agn_lf}
\end{center}
\end{figure}

The red line in Figures~\ref{fig:all_lf} to \ref{fig:agn_lf} is the double power-law prediction (Equation~\ref{equation:dpl}) for the local 90~$\mu$m luminosity function from Serjeant \& Harrison (2005), which was based on extrapolated fluxes of 15,411 local galaxies ($z<0.1$) in the PSCz catalogue (Saunders et al. 2000) derived using IRAS data:
\begin{equation}
\Phi(L)=\frac{\Phi_*}{\left[\left(\frac{L}{L_*}\right)^\beta+\left(\frac{L}{L_*}\right)^\gamma\right]}\label{equation:dpl}
\end{equation}
where $\log_{10}\Phi_*= -2.93 $\,dex$^{-1}$\,Mpc$^{-3}$,  
the characteristic break luminosity $L_*$ is given by $\log_{10}L_*=23.54$\,W\,Hz$^{-1}$\,sr$^{-1}$, 
$\beta=0.567$ and  $\gamma=2.84$.
This prediction assumes pure luminosity evolution of $(1+z)^3$. Converting to the units used in this paper (and using the more current value $H_0=72$) gives the predicted $L_*= 3.06\times10^{10}$ L$_{\odot}$.

The luminosity functions for star-forming galaxies (after excluding all those with any evidence of AGN activity) and for AGN sources are  shown separately in Figures~\ref{fig:sbg_lf} and \ref{fig:agn_lf}. In the case of AGN sources, the boxcar width is $\pm$0.2 dex due to the lower number of data points.

\begin{figure*}
 \resizebox{6.2in}{!}{\includegraphics{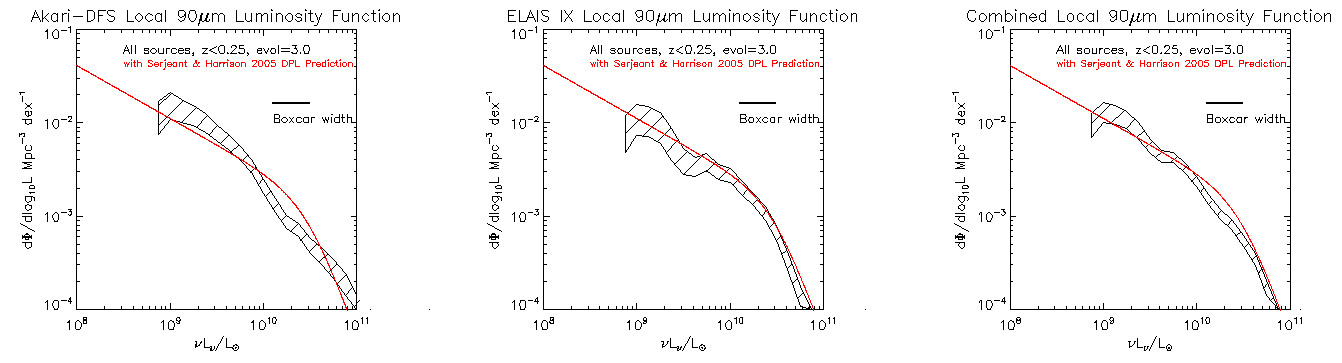}}
\caption{Comparison between 90 $\mu$m luminosity functions for (from left) ADF-S, ELAIS and combined ADF-S and ELAIS.}\label{fig:combined_lf}
\end{figure*}

\subsection{Mean of  $V/V{\tiny max}$}\label{subsec:VVmax}

The mean of  V/V{\tiny max} should be equal to 0.5 if the sample is complete, assuming a uniform space density. We found that $\langle$V/V{\tiny max}$\rangle$ = 0.478$\pm0.022$ if we assume that there is no pure luminosity evolution. However, if (1+$z$)$^3$ pure luminosity evolution is assumed,  $\langle$V/V{\tiny max}$\rangle$ falls sharply to  0.427$\pm0.023$, although the luminosity function itself is not markedly different (compare Figures~\ref{fig:all_lf} and \ref{fig:combined_lf}, left). Much of this shortfall is accounted for by the sources at 0.2$<z<$0.25: making a cut at $z$=0.2 increases  $\langle$V/V{\tiny max}$\rangle$ from 0.427 to 0.467 for the luminosity function with pure luminosity evolution of (1+$z$)$^3$. (This reduces the number of sources of sources from 130 to 119.)

We have used this value for pure luminosity evolution because earlier work has usually found pure luminosity evolution of this order, albeit with large uncertainties: for example, ELAIS IV found (1+$z$)$^{2.45\pm 0.85}$ (Serjeant et al. 2001) and ELAIS IX found  (1+$z$)$^{3.4\pm 1.0}$ (Serjeant et al. 2004), both at 90 $\mu$m, and Saunders et al. (1990) found (1+$z$)$^{3.0\pm 1.0}$ at 60 $\mu$m.

\subsection{Combined ADF-S and ELAIS Results}\label{subsec:results_combined}

The 90 $\mu$m luminosity function for the Final Analysis of the northern ELAIS fields was presented in Serjeant et al. (2004) and is reproduced in the central panel of Figure~\ref{fig:combined_lf}. This covered three Northern ELAIS fields (totalling 7.4 square degrees) to a flux limit of 70 mJy. Completeness reached ~100\% at a flux of 150 mJy and ~50\% at 70 mJy, so it was not as deep as the ADF-S survey. The ELAIS survey was specifically designed to minimise the effects of cosmic variance (Oliver et al. 2000) and so was spread over three fields, giving confidence that no cosmic variance renormalisation was necessary.

The left panel in Figure~\ref{fig:combined_lf} shows the ADF-S luminosity function recalculated assuming pure luminosity evolution of $(1+z)^3$ in order to compare and combine like-with-like with ELAIS, although as will be evident this makes only a slight difference to our luminosity function at this range ($z<0.25$) of redshifts. The right-hand panel of this Figure shows the combined (ADF-S and ELAIS) luminosity function.



\section{Discussion}\label{sec:discussion}



\begin{figure*}
 \resizebox{\hsize}{!}{\includegraphics{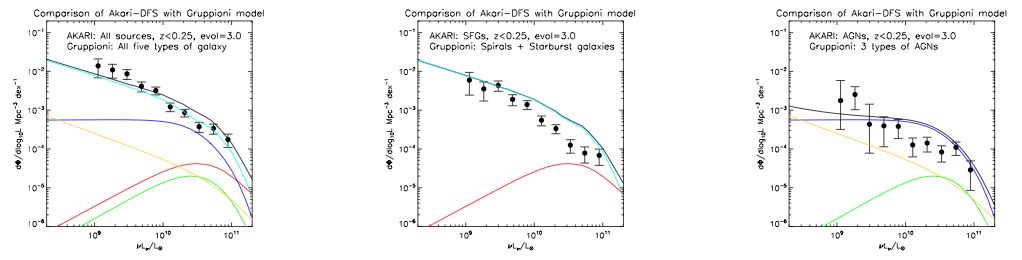}}
\caption{Comparison of the AKARI ADF-S 90 $\mu$m luminosity function with predictions from the new Gruppioni et al. (2011) model. The AKARI data points are independent in this figure (unlike the boxcar-smoothed data in Figures~\ref{fig:all_lf}-\ref{fig:combined_lf}). The plots from the Gruppioni et al. (2011) model are: black line = total, cyan= spiral galaxies, red= starburst galaxies, blue=Seyfert 2s, green=obscured AGNs and brown = AGN1 sources. The left figure shows all AKARI sources, the centre figure shows SFGs compared with the two star-forming Gruppioni classes, and the figure on the right shows AGNs.}\label{fig:gruppioni3}
\end{figure*}

The most noticeable departure from the predicted local 90 $\mu$m luminosity function of Serjeant \& Harrison (2005) described above is a deficit of galaxies from just below  L$_*$  (from $\sim10^{10} L_{\odot}$). This is particularly clear in the luminosity function where galaxies with AGN have been excluded (Figure~\ref{fig:sbg_lf}), where it is significant at about $3\sigma$. This may suggest we are sampling atypical large-scale structures, and provides strong motivation for expanding our spectroscopic area to overcome these effects. To put this in context, the comoving distance across the AAOmega field at a median redshift of $z\sim0.1$ is about 14 Mpc. The ELAIS results (Figure~\ref{fig:combined_lf}b) did not show the deficit of  L$_*$ galaxies which we have found in this study, and the combined luminosity function (Figure~\ref{fig:combined_lf}c) is close to the Serjeant \& Harrison (2005) double power-law prediction.

At the faint end of the luminosity function, there is a slight excess number density when all sources are considered compared to the Serjeant \& Harrison (2005) prediction (see Figure~\ref{fig:all_lf}). At the bright end, we find that the slope is shallower for AGNs than for SFGs (cf. Figures~\ref{fig:sbg_lf} and \ref{fig:agn_lf}) indicating an increased contribution at high luminosity from AGN sources, which we might expect given the inferences from mid-infrared spectroscopy and radiative transfer models of increased AGN bolometric fractions with increasing bolometric luminosities (see for example Genzel et al. 1998 and Genzel \& Cesarsky 2000). The relative number density of AGN sources rises from under 10\% of the total number density at the faint end to over 30\% of the number density of all sources at the bright end.

We have also compared our observations with the theoretical prediction for the 90 $\mu$m luminosity function for $0<z<0.25$ galaxies from a new backward evolution model of Gruppioni et al. (2011) which uses the methodology developed in Gruppioni et al. (2008). This comparison is shown in Figure~\ref{fig:gruppioni3} for all sources and for SFGs and AGNs separately. The Gruppioni model includes separate evolution for five types of galaxy:  spirals (with low to moderate star formation), starbursts, low-luminosity AGNs (Seyfert 2s), AGN1s, and AGN2s (obscured AGNs). The relative number densities of the populations and their evolution are constrained by the population fractions in mid-infrared selected samples taken from ISO and IRAS (e.g. La Franca et al. 2004, Matute et al. 2006; see also Manners et al. 2004), and are also constrained by early results from Herschel. In making multi-wavelength predictions for each type of object, the model uses a set of SED templates, with AGN1 being approximately flat in $\nu$L$_\nu$, while the other AGN classes have relatively stronger far-infrared excesses. Figure~\ref{fig:gruppioni3} shows the detailed predictions of this model for the different classes of galaxy, showing how the contribution from the more active galaxies is predicted to increase at higher luminosity. As with the prediction by Serjeant \& Harrison (2005), we again find a slight excess in our data at the low-luminosity end (although within our 1$\sigma$ error bars). The dip in the AKARI data around L$_*$ is not predicted by the model, but occurs for both AGNs and SBGs, suggesting it is caused by large-scale structure, as discussed above. The shallower slope in the AGN population predicted by the model is confirmed, supporting the SED assumptions of the model. Nevertheless, our AGN classification is based at optical wavelengths so we may in principle be missing AGN2s, for example, and could have misclassified AGNs and SFGs if their position on the BPT diagram does not reflect their relative bolometric contributions.

In a future paper we will extend our results by presenting additional spectroscopic redshifts in the ADF-S, covering more of the field for the mid-infrared-selected population, and also by using Herschel and other data in this field. The advent of co-ordinated wide-field far-infrared to submillimetre surveys, such as AKARI plus BLAST, or Spitzer plus Herschel, is enormously extending our ability to constrain the evolving bolometric luminosity function of galaxies.  With the additional constraints on AGN dust tori from the WISE mission covering both BLAST and Herschel survey fields it will be possible to see whether the differences we have found compared to prediction are repeated, and to improve our assessment of the relative contributions of star formation and AGNs to the evolving bolometric infrared luminosity function.

\begin{table}
\caption{The new cluster found in our redshift area of ADF-S, centred on (04 42 12.5, -53 30 46), within a radius of 45 arcminutes of this centre, and within a redshift range $0.056<z<0.061$. A near position search on the NASA Extragalactic Database (NED) yields 649 objects within this area of the sky, although virtually all (646) without known redshifts. For comparison, other searches on NED for the same sized area centred on positions offset by 2 degrees in RA and/or with reversed Dec typically yielded about 150-200 sources.}\label{table:cluster}
\begin{center}
\begin{tabular}{|ccc|}
\hline

 & & \\
~~~RA(J2000)     & ~~~Dec(J2000)  & ~~~Redshift~~~ \\ 
 
 \hline
04 40 02.7 &	-53 41 44 &    0.057\\
04 40 23.3  &	-52 59 07	&    0.059\\
04 40 29.4  &	-53 21 43  &   0.058\\
04 40 52.8 &	-53 44 18	&    0.058\\
04 40 54.6 &	-53 16 48	&    0.057\\
04 41 16.9 &	-53 47 55	&    0.057\\
04 41 24.0 &	-53 03 03	&    0.060\\
04 41 25.3 &	-52 59 53	&    0.061\\
04 41 29.3 &	-53 05 48	&    0.057\\
04 41 55.7 &	-53 11 04	&    0.061\\
04 41 57.3 &	-53 25 10	&    0.057\\
04 42 00.8 &	-53 34 22	&    0.058\\
04 42 31.4 &	-53 35 28	&    0.059\\
04 42 33.5 &	-54 09 22	&    0.060\\
04 42 51.4 &	-53 24 44	&    0.057\\
04 43 04.8 &	-54 08 01	&    0.056\\
04 43 14.4 &	-54 07 08	&    0.056\\
04 43 32.8 &	-53 29 06	&    0.057\\
04 43 42.9 &	-53 39 38	&    0.058\\
04 43 51.2 &	-53 36 27	&    0.057\\
04 44 15.4 &	-53 00 21	&    0.061\\
04 44 43.9 &	-53 55 39	&    0.059 \\
 
\hline

\end{tabular}
\end{center}
\label{default}
\end{table}

\begin{table}
\caption{ Tabulated luminosity function.}\label{table:tab}
\begin{center}
\begin{tabular}{|ccc|}
\hline
 & & \\
Luminosity bin centre & Luminosity Density &Poisson Error \\
(boxcar width $\pm$0.15 dex) &$\Phi$ &$\Delta$$\Phi$ \\
log$_{10}$(L/L$_{\odot}$)     & Mpc$^-$$^3$ dex$^-$$^1$ & Mpc$^-$$^3$ dex$^-$$^1$  \\ 

 & & \\
\hline
 &  & \\

8.768   &  	1.01x10$^-$$^2$  &    5.89x10$^-$$^3$ \\ 
8.808   &   1.27x10$^-$$^2$   &   6.41x10$^-$$^3$ \\
8.849   &  	1.27x10$^-$$^2$  &    6.41x10$^-$$^3$ \\
8.889   &  	1.48x10$^-$$^2$  &    6.75x10$^-$$^3$ \\
8.929   &  	1.63x10$^-$$^2$  &    6.90x10$^-$$^3$ \\

8.970   &  	1.80x10$^-$$^2$  &    7.12x10$^-$$^3$ \\
9.010   &  	1.35x10$^-$$^2$   &   5.31x10$^-$$^3$ \\
9.051   &  	1.16x10$^-$$^2$ &    4.55x10$^-$$^3$ \\
9.091   &  	1.00x10$^-$$^2$ &    3.91x10$^-$$^3$ \\
9.131   &  	1.15x10$^-$$^2$ &    4.05x10$^-$$^3$ \\

9.172   &  	1.08x10$^-$$^2$ &    3.60x10$^-$$^3$ \\
9.212   &  	1.11x10$^-$$^2$  &   3.45x10$^-$$^3$ \\
9.253   &  	1.04x10$^-$$^2$  &   3.06x10$^-$$^3$ \\
9.293   &  	1.04x10$^-$$^2$ &    2.86x10$^-$$^3$ \\
9.333   &  	9.72x10$^-$$^3$  &   2.56x10$^-$$^3$ \\

9.374   &  	8.42x10$^-$$^3$  &   2.18x10$^-$$^3$ \\
9.414   &  	8.28x10$^-$$^3$   &  2.08x10$^-$$^3$ \\
9.455   &  	7.29x10$^-$$^3$ &    1.84x10$^-$$^3$ \\
9.495   &  	5.37x10$^-$$^3$  &   1.43x10$^-$$^3$ \\
9.535   &  	6.30x10$^-$$^3$  &  1.49x10$^-$$^3$ \\

9.576   &  	5.76x10$^-$$^3$   &  1.30x10$^-$$^3$ \\
9.616   &  	5.10x10$^-$$^3$ &    1.19x10$^-$$^3$ \\
9.657   &  	4.95x10$^-$$^3$   &  1.01x10$^-$$^3$ \\
9.697   &  	4.56x10$^-$$^3$  &   9.29x10$^-$$^3$ \\
9.737   &  	3.92x10$^-$$^3$   &  8.04x10$^-$$^4$ \\

9.778   &  	3.79x10$^-$$^3$   &  7.62x10$^-$$^4$ \\
9.818   &  	3.13x10$^-$$^3$  &   6.32x10$^-$$^4$ \\
9.859   &  	2.85x10$^-$$^3$   &  5.71x10$^-$$^4$ \\
9.899   &  	2.28x10$^-$$^3$   &  4.83x10$^-$$^4$ \\
9.939   &  	1.60x10$^-$$^3$   &  3.56x10$^-$$^4$ \\

9.980   &  	1.47x10$^-$$^3$  &   3.19x10$^-$$^4$ \\
10.020  &  1.24x10$^-$$^3$  &  2.69x10$^-$$^4$ \\
10.061  &  1.34x10$^-$$^3$   &  2.69x10$^-$$^4$ \\
10.101  &  1.32x10$^-$$^3$  &  2.53x10$^-$$^4$ \\
10.141   &  8.97x10$^-$$^4$ &   1.89x10$^-$$^4$ \\

10.182   &  8.03x10$^-$$^4$ &   1.66x10$^-$$^4$ \\
10.222   &  8.34x10$^-$$^4$   &  1.66x10$^-$$^4$ \\
10.263   &  8.91x10$^-$$^4$  &   1.68x10$^-$$^4$ \\
10.303   &  8.15x10$^-$$^3$   &  1.56x10$^-$$^4$ \\
10.343   &  7.24x10$^-$$^4$   &  1.43x10$^-$$^4$ \\

10.384   &  5.17x10$^-$$^4$   &  1.16x10$^-$$^4$ \\
10.424   &  5.21x10$^-$$^4$  &   1.14x10$^-$$^4$ \\
10.465   &  4.38x10$^-$$^4$  &   1.03x10$^-$$^4$ \\
10.505   &  4.49x10$^-$$^4$  &   1.03x10$^-$$^4$ \\
10.546   &  3.42x10$^-$$^4$  &   8.83x10$^-$$^5$ \\

10.586   &  3.40x10$^-$$^4$  &   8.78x10$^-$$^5$ \\
10.626   &  3.11x10$^-$$^4$   &  8.33x10$^-$$^5$ \\
10.667  &   3.32x10$^-$$^4$   &  8.57x10$^-$$^5$ \\
10.707  &  2.37x10$^-$$^4$   &  7.15x10$^-$$^5$ \\
10.747  &  2.34x10$^-$$^4$   &  7.06x10$^-$$^5$ \\

10.788  &  2.28x10$^-$$^4$   &  6.88x10$^-$$^5$ \\
10.828  &  1.61x10$^-$$^4$   &  5.71x10$^-$$^5$ \\
10.869  &  1.40x10$^-$$^4$   &  5.28x10$^-$$^5$ \\

 & & \\
\hline
\end{tabular}
\end{center}
\label{default}
\end{table}

\section*{Acknowledgments}

 This research is based on observations with AKARI, a JAXA project with the participation of ESA. This work was funded in part by STFC (grant PP/D002400/1), the Royal Society (2006/R4-IJP), the Sasakawa Foundation~(3108) and KAKENHI (19540250 and 21111004). We extend our thanks to the anonymous referees whose comments were very helpful, and to Carlotta Gruppioni for providing data from her recent model.

%

\bsp           

\label{lastpage}

\end{document}